\documentclass[showpacs,showkeys,12pt,preprint,preprintnumbers,nofootinbib,groupedaddress,superscriptaddress,amsmath,amssymb]{revtex4}

\usepackage[dvips]{color}
\usepackage{graphicx}
\usepackage{epsfig}    
\usepackage{dcolumn}
\usepackage{bm}
\usepackage{here}

\def\Journal#1#2#3#4{{#1} {\bf #2}, #3 (#4)}

\def\NPB{{ Nucl. Phys.} B}
\def\PLB{{ Phys. Lett.}  B}

\def\PRD{{ Phys. Rev.} D}

\allowdisplaybreaks[4]

\newcommand{\gae}{\stackrel{>}{\sim}}

\begin{document}

\title{
Higgs boson pair production at a photon-photon collision\\
in the two Higgs doublet model
}

\author{Eri Asakawa}
\email{eri@post.kek.jp}
\affiliation{Institute of Physics, Meiji Gakuin University, Yokohama 244-8539, Japan}
\author{Daisuke~Harada}
\email{dharada@post.kek.jp}   
\affiliation{Theory Group, Institute of Particle and Nuclear Studies,
KEK, 1-1 Oho, Tsukuba, Ibaraki 305-0801, Japan}
\affiliation{Department of Particle and Nuclear Physics, the Graduate
University for Advanced Studies (Sokendai), 1-1 Oho, Tsukuba, Ibaraki 305-0801, Japan}
\author{Shinya~Kanemura}
\email{kanemu@sci.u-toyama.ac.jp}
\affiliation{Department of Physics, University of Toyama, 3190 Gofuku, Toyama 930-8555, Japan}
\author{Yasuhiro~Okada}
\email{yasuhiro.okada@kek.jp}   
\affiliation{Theory Group, Institute of Particle and Nuclear Studies,
KEK, 1-1 Oho, Tsukuba, Ibaraki 305-0801, Japan}
\affiliation{Department of Particle and Nuclear Physics, the Graduate
University for Advanced Studies (Sokendai), 1-1 Oho, Tsukuba, Ibaraki 305-0801, Japan}
\author{Koji~Tsumura}
\email{ktsumura@ictp.it}   
\affiliation{International Centre for Theoretical Physics, Strada
Costiera 11, 34014 Trieste, Itary}

\preprint{KEK-TH-1271, UT-HET 014, IC/2008/59}
\pacs{12.60.Fr, 14.70.Bh, 14.80.Cp}
\keywords{Higgs self-coupling, photon collider, new physics}

\begin{abstract}

We calculate the cross section of Higgs boson pair production at a
photon collider in the two Higgs doublet model. 
We focus on the scenario in which the lightest CP even Higgs boson ($h$)
has the standard model like couplings to the gauge bosons. 
We take into account the one-loop correction to the $hhh$ coupling as  
well as additional one-loop diagrams due to charged Higgs bosons to 
the $\gamma\gamma\to hh$ helicity amplitudes. 
It is found that the full cross section can be enhanced by both these effects
  to a considerable level.
We discuss the impact of these corrections on the $hhh$ coupling
measurement at the photon collider. 
\end{abstract}

\maketitle


The Higgs sector is the last unknown part of the standard model (SM)
for elementary particles. 
Discovery of the Higgs boson and the measurement of its properties
at current and future experiments are crucial to establish our basic
picture for spontaneous electroweak symmetry breaking (EWSB)
and the mechanism of particle mass generation. 
The Higgs mechanism would be experimentally tested after the discovery of a
new scalar particle by measuring its mass and the coupling to the weak gauge
bosons. The mass generation mechanism for quarks and charged leptons
via the Yukawa interaction is also clarified by the precise
determination of both the fermion masses and the Yukawa
coupling constants.
If the deviation from the SM relation between the mass and
the coupling is found, it can be regarded as an evidence of new physics
beyond the SM. 
The nature of EWSB can be revealed through the experimental
reconstruction of the Higgs potential, for which the measurement
of the Higgs self-coupling is essential\cite{hhh-lhc,hhh-lhc-ilc,eehhZ1,eehhZ2,eehhZ3}.
The structure of the Higgs potential depends on the scenario of
new physics beyond the SM\cite{hhh-thdm1,hhh-thdm2}, so that the experimental determination of
the triple Higgs boson coupling can be a probe of each new physics
scenario.
Furthermore, the property of the Higgs potential would be 
directly related to the aspect of the electroweak phase transition
in the early Universe, which could have impact on the problem of
the electroweak baryogenesis\cite{ewbg}. 

It is known that the measurement of the triple Higgs boson coupling is rather challenging
at the CERN Large Hadron Collider (LHC), requiring huge luminosity.  A study has shown that at the
SLHC with the luminosity of 3000 fb$^{-1}$, expected accuracy would be
about $20$-$30$\% for the mass ($m_h$) of the Higgs boson ($h$) to be around 170 GeV\cite{hhh-lhc,hhh-lhc-ilc}.
At the international linear collider (ILC), the main processes for the
$hhh$ measurement are the double Higgs boson production mechanisms via
the Higgs-stlahlung and the W-boson fusion\cite{eehhZ1,eehhZ2}.  If the collider energy is
lower than 1 TeV, the double Higgs strahlung process $e^+e^- \to Z hh$
is important for a light Higgs boson with the mass of 120-140 GeV,
while for higher energies the W-boson fusion process $e^+e^- \to hh \nu \bar \nu$ becomes dominant
due to its $t$-channel nature\cite{eehhZ3}. 
Sensitivity to the $hhh$ coupling in these processes becomes rapidly worse for greater Higgs boson masses.
In particular, for the intermediate mass range (140 GeV $< m_h <$ 200
GeV), it has not yet been known how accurately the $hhh$ coupling can be
measured by the electron-positron collision.

The photon collider is an option of the ILC. The possibility of
measuring the $hhh$ coupling via the process of $\gamma\gamma\to hh$ has
been discussed in Ref.~\cite{jikia}, where the cross section has been
calculated at the one-loop level, and the dependence on the triple Higgs
boson coupling constant is studied.
In Ref.~\cite{belusevic} the statistical sensitivity to the $hhh$
coupling constant has been studied especially for a light Higgs boson
mass in relatively low energy collisions.
Recently, these analyses have been extended for wider regions of the Higgs
boson masses and the collider energies. It has been found that 
when the collision energy is limited to be lower than 500-600 GeV
the statistical sensitivity to the $hhh$ coupling can be better for the
process in the $\gamma\gamma$ collision than
that in the electron-positron collision for the Higgs boson with the
mass of 160 GeV \cite{tilc08}. 

Unlike the double Higgs production processes $e^+e^- \to Zhh$ and 
$e^+e^-\to hh \nu\bar\nu$ in $e^+e^-$ collisions,
$\gamma\gamma \to hh$ is an one-loop induced process. 
When the origin of the shift in the $hhh$ coupling would be due to 
one-loop corrections by new particles, it may also
affect the amplitude of $\gamma\gamma \to hh$ directly through 
additional one-particle-irreducible (1PI) one-loop diagrams of
$\gamma\gamma h$ and $\gamma\gamma hh$ vertices. 

In this letter, we consider the new particle effect on the
$\gamma\gamma \to hh$ cross sections in the two Higgs doublet model
(THDM), in which additional CP-even, CP-odd and charged Higgs bosons appear.
It is known that a non-decoupling one-loop effect due to these extra Higgs
bosons can enhance the $hhh$ coupling constant by ${\cal O}(100)$ \%\cite{hhh-thdm1}.
In the $\gamma\gamma\to hh$ helicity amplitudes, there are additional
one-loop diagrams by the charged Higgs boson loop to the ordinary
SM diagrams (the W-boson loop and the top quark loop).
We find that both the charged Higgs boson loop contribution to the
$\gamma\gamma\to hh$ amplitudes and the non-decoupling effect on the $hhh$
coupling can enhance the cross section from its SM value significantly.  
We consider how the new contribution to the cross section of $\gamma\gamma\to hh$ would affect the
measurement of the triple Higgs boson coupling at a $\gamma\gamma$ collider.


In order to examine the new physics effect on $\gamma\gamma \to hh$,
we calculate the helicity amplitudes in the THDM.
We impose a discrete symmetry to the model to avoid flavor changing neutral current in a natural
way\cite{Glashow-Weinberg}. 
The Higgs potential is then given by
\begin{eqnarray}
 V_{\rm THDM}&=& \mu_1^2 |\Phi_1|^2+\mu_2^2 |\Phi_2|^2-(\mu_3^2
  \Phi_1^\dagger \Phi_2 + {\rm h.c.})\nonumber\\
&&  + \lambda_1 |\Phi_1|^4
  + \lambda_2 |\Phi_2|^4
 + \lambda_3 |\Phi_1|^2|\Phi_2|^2
  + \lambda_4 |\Phi_1^\dagger \Phi_2|^2
  + \frac{\lambda_5}{2} \left\{(\Phi_1^\dagger \Phi_2)^2 + {\rm h.c.}
                        \right\}, 
\end{eqnarray}
where $\Phi_1$ and $\Phi_2$ are two Higgs doublets with hypercharge
$+1/2$. We here include the soft breaking term for the discrete symmetry by
the parameter $\mu_3^2$.
In general, $\mu_3^2$ and $\lambda_5$ are complex, but
we here take them to be real for simplicity.
We parameterize the doublet fields as
 \begin{eqnarray}
  \Phi_i = \left[\begin{array}{c}
            \omega_i^+ \\ \frac{1}{\sqrt{2}}(v_i+h_i + i z_i)
            \end{array}
   \right], \hspace{4mm} (i=1,2),  
  \end{eqnarray}
  where vacuum expectation values (VEVs) $v_1$ and $v_2$ satisfy
  $v_1^2+v_2^2 = v^2 \simeq (246 \hspace{2mm} {\rm GeV})^2$. 
  The mass matrices can be diagonalized by introducing the mixing angles
  $\alpha$ and $\beta$, where $\alpha$ diagonalizes the mass matrix of
  the CP-even neutral bosons, and $\tan\beta=v_2/v_1$.
  Consequently, we have two CP even ($h$ and $H$), a CP-odd ($A$) and
  a pair of charged ($H^\pm$) bosons.
  We define $\alpha$ such that $h$ is the SM-like Higgs boson when $\sin(\beta-\alpha)=1$.
  We do not specify the type of Yukawa interactions\cite{Barger},
  because it does not much affect the following discussions. 

  Throughout this letter, we concentrate on the case with so called the SM-like limit [$\sin(\beta-\alpha)=1$],
  where the lightest Higgs boson $h$ has the same tree-level  couplings as the SM
  Higgs boson, and the other bosons do not couple to gauge bosons and behave just as
  extra scalar bosons. 
  In this limit, the masses of the Higgs bosons are\footnote{For the case without the SM-like limit, see Ref.~\cite{hhh-thdm2} for example.} 
  \begin{eqnarray}
    m_h^2 &=& \{\lambda_1 \cos^4\beta+\lambda_2 \sin^4\beta + 2
     (\lambda_3+\lambda_4+\lambda_5) \cos^2\beta\sin^2\beta\} v^2,\\
    m_H^2 &=& M^2 +
     \frac{1}{8}\left\{\lambda_1+\lambda_2-2(\lambda_3+\lambda_4+\lambda_5)\right\}(1-\cos
      4\beta)v^2,\\
    m_A^2 &=& M^2 - \lambda_5 v^2,\\
    m_{H^\pm}^2 &=& M^2 - \frac{\lambda_4+\lambda_5}{2}v^2, 
  \end{eqnarray}
where $M (= |\mu_3|/\sqrt{\sin\beta\cos\beta})$ represents the soft
breaking scale for the discrete symmetry, and determines the decoupling property of the extra Higgs
bosons. 
When $M \sim 0$, the extra Higgs bosons $H$, $A$ and $H^\pm$ receive their masses from
the VEV, so that the masses are proportional to $\lambda_i$.
Large masses cause significant non-decoupling effect in the radiative correction to the $hhh$ coupling.
On the other hand, when $M \gg v$
the masses are determined by $M$. In this case, the quantum effect
decouples for $M \to \infty$.
  
There are several important constraints on the THDM parameters from the
data. The LEP direct search results give the lower bounds  $m_h > 114$ GeV in the SM-like limit and  
$m_H^{}$, $m_A^{}$, $m_{H^\pm}^{} \gae$ 80-90 GeV\cite{pdg}. 
In addition, the rho parameter data at the LEP requires the approximate custodial symmetry
in the Higgs potential. This implies that
$m_{H^\pm}^{} \simeq m_A^{}$ or $\sin(\beta-\alpha) \simeq 1$ and $m_{H^\pm}
\simeq m_H^{}$. 
The Higgs potential is also constrained from the tree level unitarity\cite{pu,pu1},
the triviality and vacuum stability\cite{vs}, in particular for the case
where the non-decoupling effect is important as in the discussion here.
For $M \sim 0$, masses of the extra Higgs bosons $H$, $A$ and $H^\pm$
are bounded from above by about $500$ GeV for $\tan\beta=1$, when they are degenerated\cite{pu}.
With non-zero $M$, these bounds are relaxed depending on the value of $M$.
The constraint from $b\to s \gamma$  gives a lower bound on the mass of $H^\pm$
depending on the type of Yukawa interaction; i.e.,  in Model
II\cite{Barger}, $m_{H^\pm}^{} > 295$ GeV ($95$\% CL)\cite{bsg}.
Recent data for $B\to\tau \nu$ can also give a constraint on the charged Higgs
mass especially for large values of $\tan\beta$ in Model II\cite{btn,btn2}.
In the following analysis, we do not include these constraints from 
B-physics because we do not specify type of Yukawa interactions.

In the THDM with $\sin(\beta-\alpha)=1$, the one-loop helicity amplitudes for the initial photon helicities
$\ell_1$ and $\ell_2$ ($\ell_i=+1$ or $-1$) are given as 
\begin{eqnarray}
 {\cal M}_{\rm THDM}^{\rm 1-loop}(\ell_1,\ell_2)={\cal M}(\ell_1,\ell_2,\lambda_{hhh})+\Delta{\cal
  M}(\ell_1,\ell_2,\lambda_{hhh}),  \label{eq-amp1loop}
\end{eqnarray}
where $\lambda_{hhh} = -3m_h^2/v$, ${\cal M}(\ell_1,\ell_2,\lambda_{hhh})$ is the SM amplitude given in
Ref.~\cite{jikia}, and  $\Delta{\cal M}(\ell_1,\ell_2,\lambda_{hhh})$ represents additional
one-loop contributions from the charged Higgs boson loop to the
$\gamma\gamma\to hh$ cross section.
We note that $\lambda_{hhh}$ has the same form as in the SM when $\sin(\beta-\alpha)=1$.
Due to the parity we have 
${\cal M}_{\rm THDM}(\ell_1,\ell_2)={\cal M}_{\rm THDM}(-\ell_1,-\ell_2)$,
so that there are independent two helicity amplitudes.

\begin{figure}[t]
\begin{center}
\includegraphics[width=12cm]{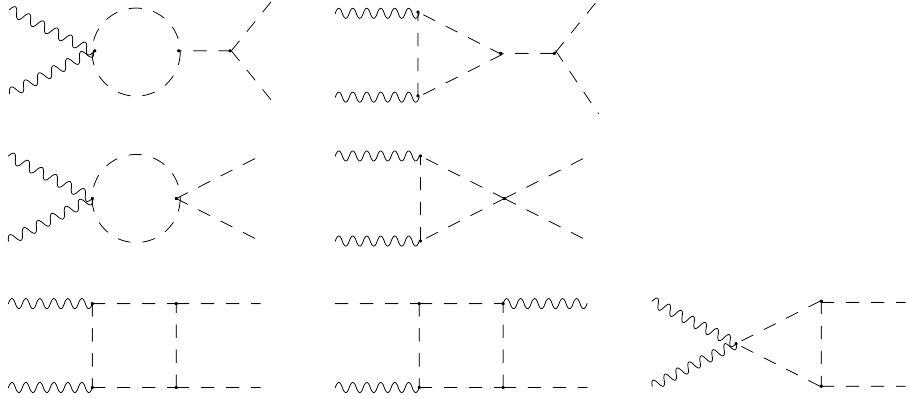}
\caption{%
Feynman diagrams for charged Higgs boson contributions to $\gamma\gamma
 \to hh$. Wavy lines represent photons, dotted lines in loops represent
 charged Higgs bosons $H^\pm$, and other dotted lines are the neutral Higgs
 bosons.
}
\label{figure}
\end{center}
\end{figure}
The Feynman diagrams which contribute to $\Delta {\cal M}$ are shown in
Fig.~1. $\Delta {\cal M}$ is given for each helicity set for
$\sin(\beta-\alpha)\simeq 1$ as  
\begin{eqnarray}
\alpha_W \Delta {\cal M}(+,+,\lambda_{hhh}) &=& 
  \frac{12\lambda_{hH^+H^-}^{} \lambda_{hhh}}{\hat{s}-m_h^2}
   \left\{ C_{24}(\hat{s})
- \frac{1}{4} B_0(\hat{s},m_{H^\pm}^{},m_{H^\pm}^{})
   \right\} \nonumber\\
&&+ 4\lambda_{hhH^+H^-}^{} B_0(\hat{s},m_{H^\pm}^{},m_{H^\pm}^{}) 
- (\lambda_{hH^+H^-}^{})^2
  \tilde{C}_0(\hat{s})
- 4\lambda_{hhH^+H^-}^{}
  C_{24}(\hat{s}) \nonumber\\
 &&
+  (\lambda_{hH^+H^-}^{})^2\left\{
\frac{}{}  \left(
 D_{27}^{1234}+D_{27}^{1243}+D_{27}^{2134}+D_{27}^{2143}
  \right) \right. \nonumber\\
&&\left.  -\frac{1}{2\hat{s}} \left(\hat{t}\hat{u}-m_h^4 \right)
  \left(
 D_{23}^{1234}+D_{23}^{1243}+D_{23}^{2134}+D_{23}^{2143}
  \right)\right\},
\end{eqnarray}
and 
\begin{eqnarray}
\alpha_W \Delta {\cal M}(+,-,\lambda_{hhh}) &=& 
 - (\lambda_{hH^+H^-}^{})^2
 \frac{1}{2\hat{s}} \left(\hat{t}\hat{u}-m_h^4 \right)
  \left(
 D_{23}^{1234}+D_{23}^{1243}+D_{23}^{2134}+D_{23}^{2143}
  \right), 
\end{eqnarray}
where $\hat s$, $\hat t$ and $\hat u$ are ordinary Mandelstam variables
for the sub processes, and  
$C_{24}(\hat{s})=C_{24}(0,0,\hat{s},m_{H^\pm}^{},m_{H^\pm}^{},m_{H^\pm}^{})$, 
$\tilde{C}_{0}(\hat{s})=C_0(m_h^2,m_h^2,\hat{s},m_{H^\pm}^{},m_{H^\pm}^{},m_{H^\pm}^{})$,
and 
$D_{ab}^{ijkl}=D_{ab}(p_i^2,p_j^2,p_k^2,p_l^2,m_{H^\pm}^{},m_{H^\pm}^{},m_{H^\pm}^{},m_{H^\pm}^{})$. 
Here we employ the Passarino-Veltman formalism in Ref.~\cite{pv}.
We take the same normalization for these amplitudes as in Ref.~\cite{jikia}.
We note that $\Delta {\cal M}(+,-,\lambda_{hhh})$ is independent of
$\lambda_{hhh}$ because of no ${\hat s}$-channel diagram contribution.  
The scalar coupling constants $\lambda_{hH^+H^-}^{}$ and $\lambda_{hhH^+H^-}^{}$ are defined by 
\begin{eqnarray}
 \lambda_{hH^+H^-}^{} =  2 \lambda_{hhH^+H^-}^{} = -
  \left(\frac{m_h^2}{v}+2\frac{m_{H^\pm}^2-M^2}{v}\right).
 \end{eqnarray}
The relative sign between ${\cal M}(\ell_1,\ell_2,\lambda_{hhh})$
and $\Delta {\cal M}(\ell_1,\ell_2,\lambda_{hhh})$ has been
checked to be consistent with the results for the effective Lagrangian in Eq.~(19)
in Ref.~\cite{vainshtein} in the large mass limit for inner particles. 

In Eq.~(\ref{eq-amp1loop}), $\lambda_{hhh}$ is the tree level coupling
constant. It is known that in the THDM $\lambda_{hhh}$ can be changed
by the one-loop contribution of extra Higgs bosons due to the
non-decoupling effect (when $M \sim 0$).
In the following analysis, we include such an effect on the cross
sections replacing $\lambda_{hhh}$ by 
the effective coupling $\Gamma_{hhh}^{\rm THDM}(\hat{s},m_h^2,m_h^2)$,
which is evaluated at the one-loop level as\cite{hhh-thdm1}
\begin{eqnarray}
\Gamma_{hhh}^{\rm THDM}(\hat{s},m_h^2,m_h^2) \simeq - \frac{3 m_h^2}{v} \left[ 1 + \sum_{\Phi=H,A,H^+,H^-}
                               \frac{m_{\Phi}^4}{12 \pi^2 v^2 m_h^2}
                              \left(1-\frac{M^2}{m_\Phi^2}\right)^3 -
                              \frac{N_c m_t^4}{3\pi^2 v^2 m_h^2} \right]. \label{hhh-2hdm}
\end{eqnarray}
As a striking feature, there are quartic power contributions of the
masses of extra Higgs bosons which is divided by $v^2 m_h^2$, when $M
\sim 0$.
Thus, the large mass of the extra Higgs boson ($H$, $A$,
$H^\pm$) with  the lighter  SM-like Higgs boson $h$ would cause large quantum
corrections to the ${hhh}$ coupling, which amount to 50-100\%.
This effect can be regarded as the leading two loop contribution to
$\gamma\gamma\to hh$ in our analysis.
The exact one-loop formula for $\Gamma_{hhh}^{\rm THDM}$ is given in
Ref.~\cite{hhh-thdm2}, which has been used in our actual numerical analysis.

Finally, the cross section for the each subprocess is given by\footnote{
The right hand side of Eq.~(\ref{eq-sc}) is different from the formula
in Ref.~\cite{jikia} by factor 1/2, but Eq.~(\ref{eq-sc}) reproduces
figures shown in Ref.~\cite{jikia}.}  
\begin{eqnarray}
  \frac{d\hat{\sigma}(\ell_1,\ell_2)}{d\hat{t}}=\frac{\alpha^2\alpha_W^{2}}{32\pi
  \hat{s}^2} |{\cal M}_{\rm THDM}(\ell_1,\ell_2)|^2, \label{eq-sc}
\end{eqnarray}
where
 ${\cal M}_{\rm THDM}^{\rm 2-loop}(\ell_1,\ell_2)$ is defined by 
\begin{eqnarray}
 {\cal M}_{\rm THDM}^{\rm 2-loop}(\ell_1,\ell_2)={\cal
  M}(\ell_1,\ell_2,\Gamma_{hhh}^{\rm THDM})+\Delta{\cal
  M}(\ell_1,\ell_2,\Gamma_{hhh}^{\rm THDM}).  \label{eq-amp2loop}
\end{eqnarray}

We comment on the consistency of our perturbation calculation.  
One might think that the inclusion of the one-loop corrected $hhh$ vertex
function $\Gamma_{hhh}^{\rm THDM}$ in the
calculation of the cross section $\gamma\gamma \to hh$ would be
inconsistent unless we also take all the other two loop contributions
into account.
Our calculation can be justified in the following sense.
First of all, 
$\Gamma_{hhh}^{\rm THDM}$ is a gauge invariant subset.
Second, it can be seen from Eq.~(\ref{hhh-2hdm}) that the deviation from the SM value
$\Delta \Gamma_{hhh}^{\rm THDM}/\Gamma_{hhh}^{\rm SM}$
$(\equiv \Gamma_{hhh}^{\rm THDM}/\Gamma_{hhh}^{\rm SM} -1)$, where
   $\Gamma_{hhh}^{\rm SM}$ is one-loop vertex function of $hhh$ in the
   SM given in Ref.~\cite{hhh-thdm2}, 
can be
of ${\cal O}(1)$ for the case of $M^2$, $m_h^2$ $\ll$ $m_{\Phi}^{2}$,  
whereas the contributions from the other two loop diagrams do not contain the 
factor $m_\Phi^2/m_h^2$, and thus relatively unimportant for $m_\Phi^2 \gg m_h^2$.
Therefore, we can safely neglect these effects as compared to
the non-decoupling loop effect in the $hhh$ coupling.
The details are shown in Appendix. 


\begin{figure}[t]
\begin{center}
\includegraphics[width=8cm,angle=-90]{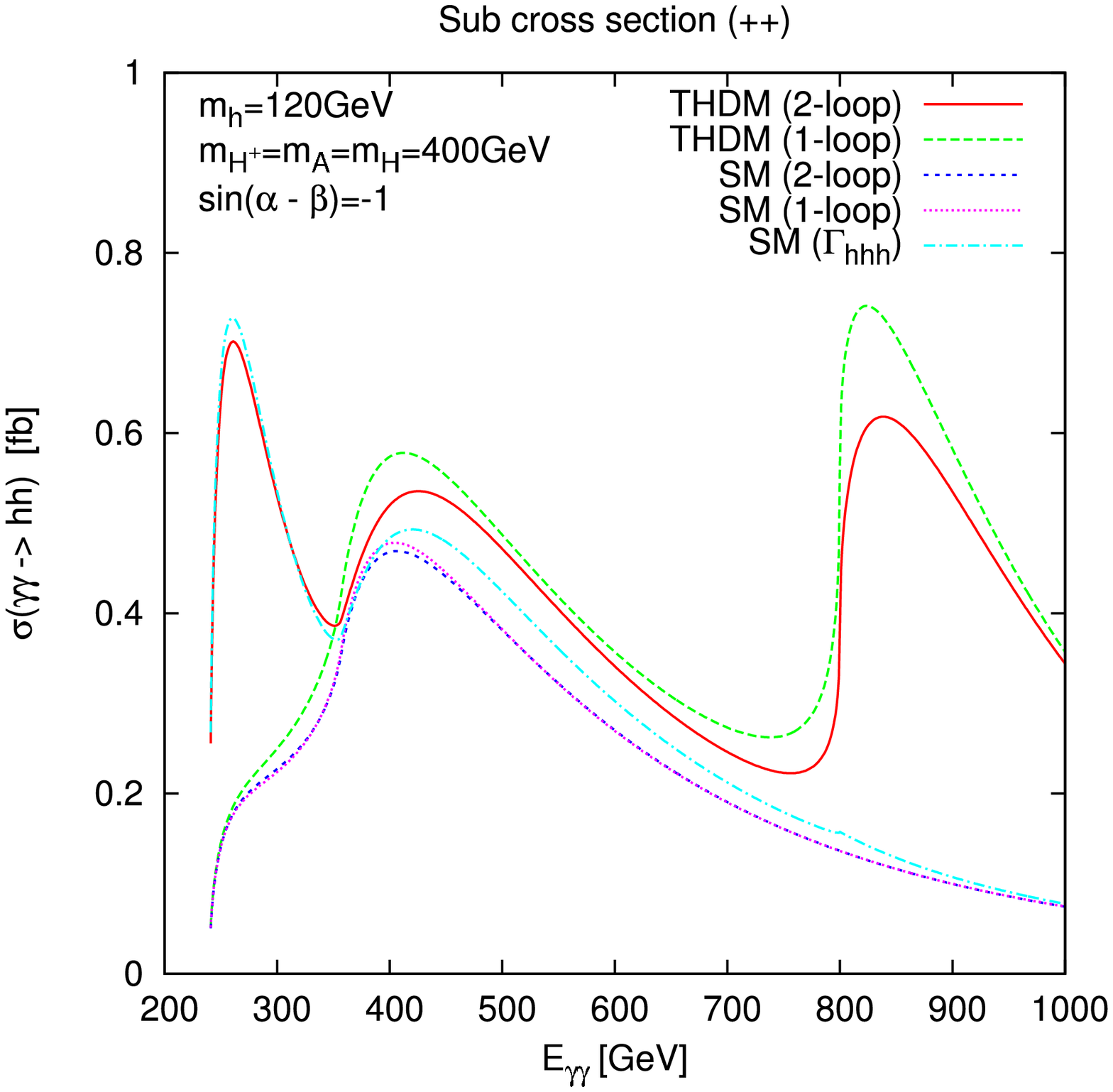}
\includegraphics[width=8cm,angle=-90]{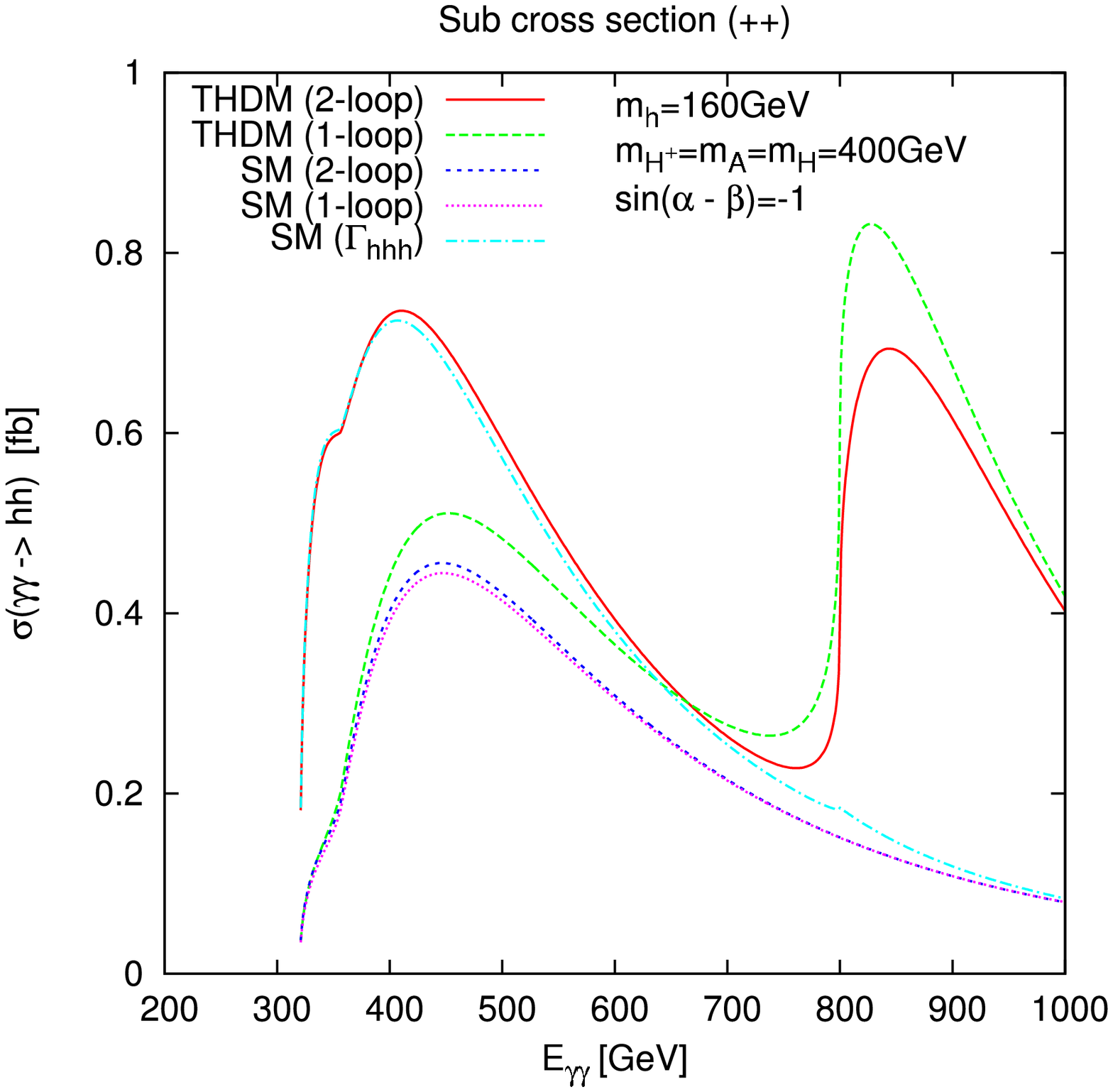}
\caption{%
 The cross section $\hat{\sigma}(+,+)$
 for the sub process $\gamma\gamma\to hh$ with the photon helicity set 
  $(+,+)$ as a function of the collision energy $E_{\gamma\gamma}$.
 In the left [right] figure the parameters are taken to be $m_h=120$
 [$160$] GeV for $m_{\Phi}^{} (\equiv m_H^{}=m_A^{}=m_{H^\pm}^{}) =400$
 GeV,  
 $\sin(\beta-\alpha)=1$, $\tan\beta=1$ and $M=0$.
}
\label{figure2}
\end{center}
\end{figure}
In Fig.~\ref{figure2}, the cross sections of
$\gamma\gamma\to hh$ for the helicity set $(+,+)$ are shown as
a function of the photon-photon collision energy $E_{\gamma\gamma}$.
In the left [right] figure, parameters are chosen to be $m_h=120$ GeV
[$m_h=160$ GeV], $\sin(\beta-\alpha)=1$, $\tan\beta=1$, $M=0$ and
$m_H^{}=m_A^{}=m_{H^\pm}^{}=400$ GeV.
In this case,
$\Delta \Gamma_{hhh}^{\rm THDM}/\Gamma_{hhh}^{\rm SM}$ amounts to about 120\% for
$m_h=120$ GeV (80\% for $m_h=160$ GeV)\cite{hhh-thdm1}
\footnote{The results of $\Delta \Gamma_{hhh}^{\rm
THDM}/\Gamma_{hhh}^{\rm SM}$ with
$M \neq 0$ are given in Ref.~\cite{hhh-thdm1}.}. 
The five curves in each figure correspond to the following cases,
\begin{itemize}
\item[(a)]
     THDM 2-loop: the cross section in the THDM with additional one-loop corrections to the
$hhh$ vertex, $\Gamma_{hhh}^{\rm THDM}$;
i.e., the contribution from ${\cal M}_{\rm THDM}^{\rm 2-loop}(+,+)$ in Eq.~(\ref{eq-amp2loop}). 

\item[(b)]
 THDM 1-loop: the cross section  in the THDM with the tree level $hhh$ coupling constant $\lambda_{hhh}$; i.e., the contribution from ${\cal
M}_{\rm THDM}^{\rm 1-loop}(+,+)$ in Eq.~(\ref{eq-amp1loop}).  

\item[(c)]
SM 2-loop: the cross section in the SM with additional top loop correction to the $hhh$
coupling $\Gamma_{hhh}^{\rm SM}$ given in Ref.~\cite{hhh-thdm2}.

\item[(d)]
SM 1-loop: the cross section in the SM with the
tree level $hhh$ coupling constant $\lambda_{hhh}^{\rm SM}$ ($=\lambda_{hhh}$ for $\sin(\beta-\alpha)=1$).  

\item[(e)]
 For comparison, we also show the result which corresponds to
the SM 1-loop result with the effective $hhh$ coupling 
$\Gamma^{\rm THDM}_{hhh}$.
\end{itemize}
In the left figure, there are three peaks in the 2-loop THDM cross section.
The one at the lowest $E_{\gamma\gamma}$ is the peak just above  the
threshold of $hh$ production. 
There the cross section is by about factor three enhanced as
compared to the SM prediction due to the effect
of $\Delta \Gamma_{hhh}^{\rm THDM}/\Gamma_{hhh}^{\rm SM}$  ($\sim 120$\%) because of the
dominance of the pole diagrams in $\gamma\gamma\to hh$.
 The second peak at around $E_{\gamma\gamma} \sim$ 400 GeV comes from the
 top quark loop contribution which is enhanced by the threshold
 of  top pair production.
 Around this point, the 2-loop THDM cross section in the case (a) can
 be well described by that in the case (e). 
 For $E_{\gamma\gamma} \sim 400$-$600$ GeV,  
 the cross section in the THDM 2-loop result deviates from the SM value 
 due to both the charged Higgs loop effect in $\Delta {\cal M}$ and
 the effect of $\Delta \Gamma_{hhh}^{\rm THDM}/\Gamma_{hhh}^{\rm SM}$.
 The third peak at around $E_{\gamma\gamma} \sim$ 850 GeV is the
 threshold enhancement of the charged Higgs boson loop in $\Delta {\cal M}$, where 
 the real production of charged Higgs bosons occurs.
 The contribution from the non-pole one-loop diagrams is dominant.
In the right figure, we can see two peaks around $E_{\gamma\gamma} \sim$
350-400 GeV and 850 GeV.
At the first peak, the contribution from the pole diagrams is dominant
so that the cross section is largely enhanced by the effect of $\Delta
\Gamma_{hhh}^{\rm THDM}/\Gamma_{hhh}^{\rm SM}$ by
several times 100\% for $E_{\gamma\gamma} \sim 350$ GeV.
It also amounts to about 80\% for $E_{\gamma\gamma} \sim 400$ GeV.
For $E_{\gamma\gamma} < 600$-700 GeV, the result in the case (e) gives a
good description of that in the case (a). 
The second peak is due to the threshold
effect of the real $H^+H^-$ production as in the left figure.

The full cross section of $e^-e^- \to \gamma\gamma \to hh$ is given from
the sub cross sections by convoluting the photon
luminosity spectrum\cite{jikia}: 
\begin{eqnarray}
  d\sigma=\int_{4m_h^2/s}^{y_m^2}
  d\tau \frac{dL_{\gamma\gamma}}{d\tau}
  \left\{
  \frac{1+\xi_1\xi_2}{2}d\hat{\sigma}(+,+) +
  \frac{1-\xi_1\xi_2}{2}d\hat{\sigma}(+,-) 
  \right\}, 
 \end{eqnarray}
where $\sqrt{s}$ is the centre-of-mass energy of the $e^-e^-$ system,
and   
\begin{eqnarray}
  \frac{dL_{\gamma\gamma}}{d\tau}
  =\int_{\tau/y_m}^{y_m} \frac{dy}{y} f_\gamma(x,y) f_\gamma(x,\tau/y), 
 \end{eqnarray}
where $\tau=\hat{s}/s$, $y=E_\gamma/E_b$ with $E_\gamma$ and $E_b$ being the energy of
photon and electron beams respectively, and $y_m=x/(1+x)$ with $x=4E_b
\omega_0/m_e^2$ where $\omega_0$ is the laser photon energy and $m_e$ is
the electron mass. In our study, we set $x=4.8$. 
The photon momentum distribution function $f_\gamma(x,y)$ and mean
helicities of the two photon beams $\xi_i$ ($i=1,2$) are given in Ref.~\cite{psf}.
\begin{figure}[t]
\begin{center}
\includegraphics[width=8cm,angle=-90]{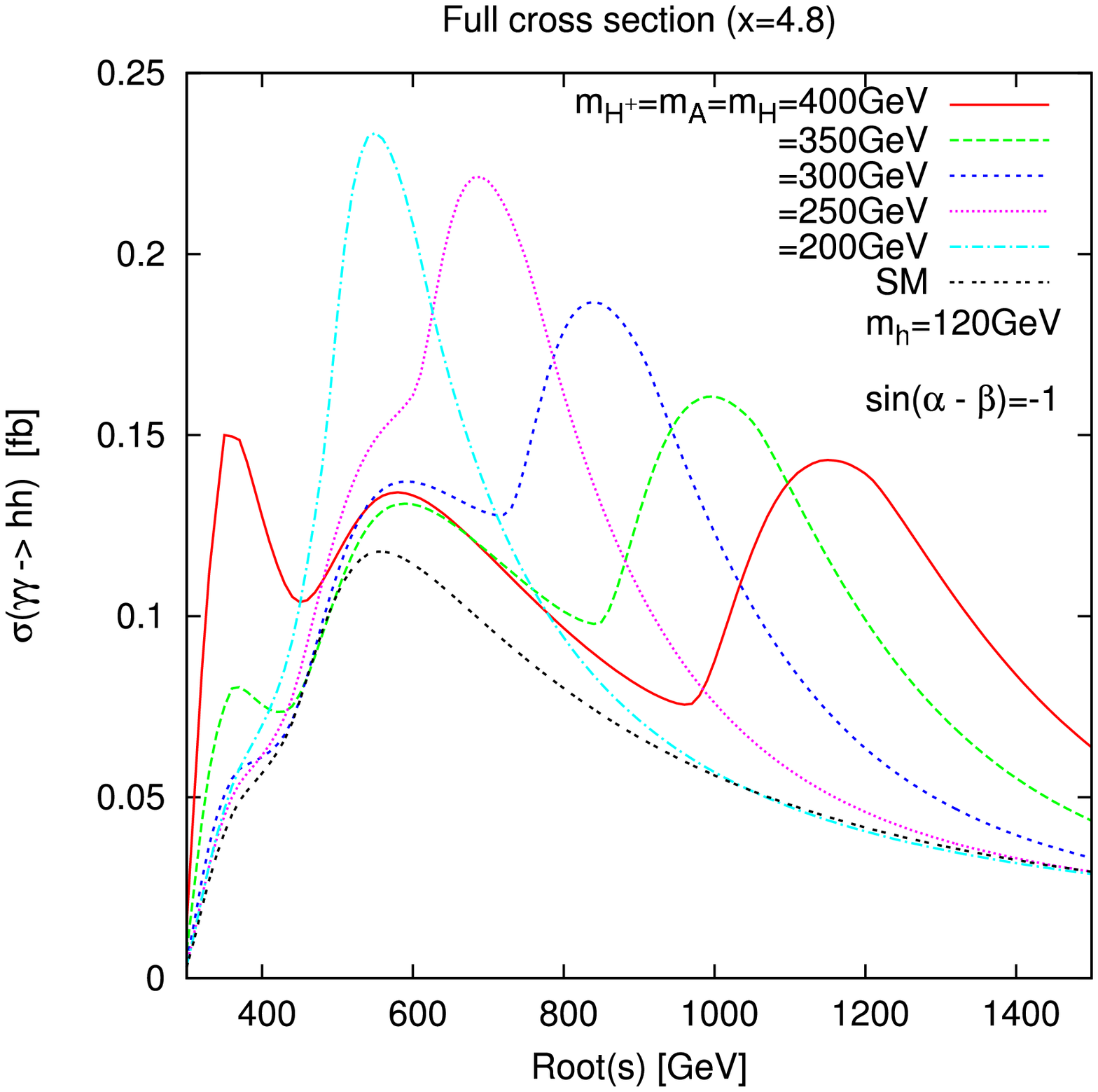}
\includegraphics[width=8cm,angle=-90]{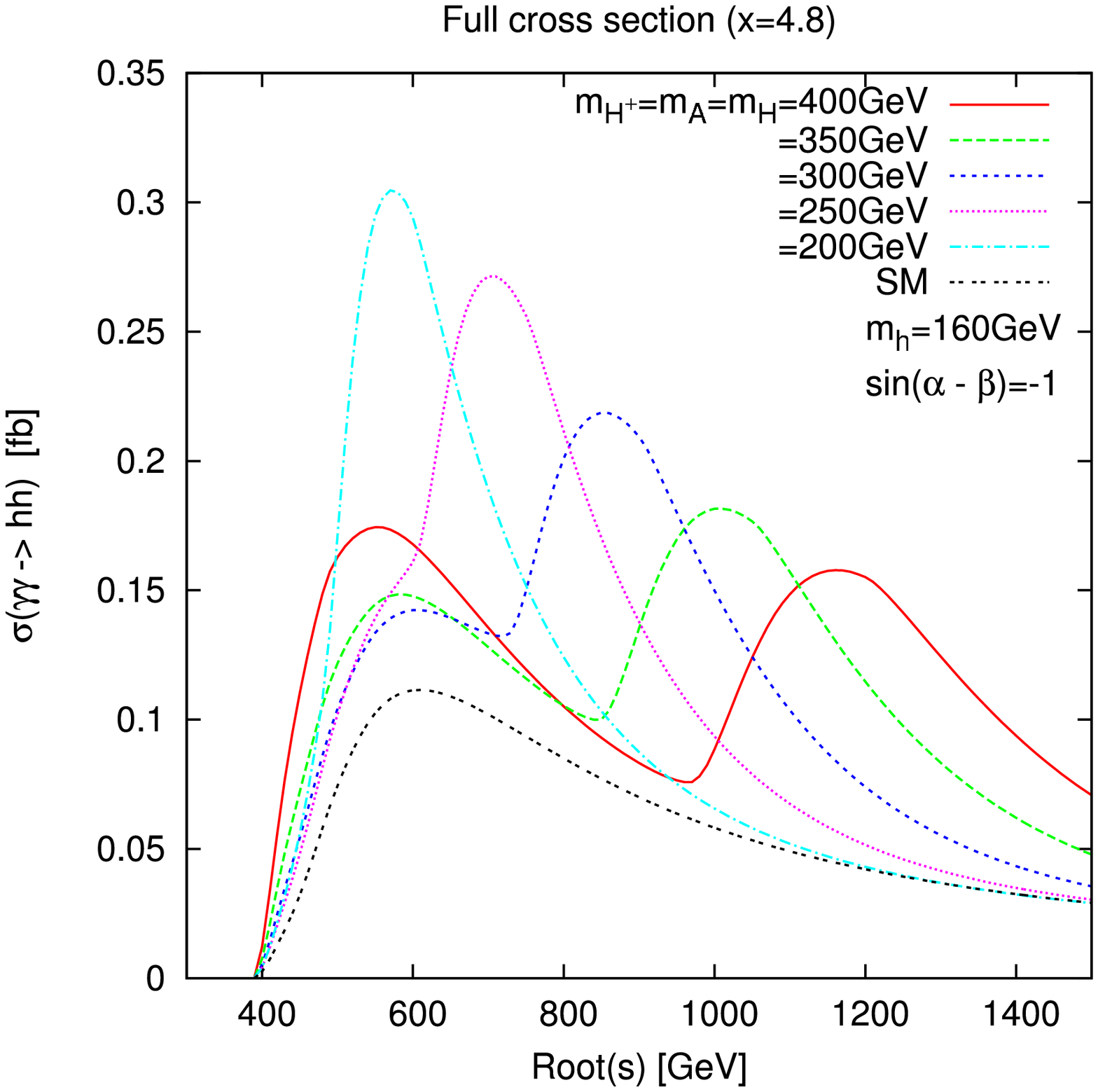}
\caption{%
 The full cross section of $e^-e^- \to \gamma\gamma \to hh$ as a
 function of $\sqrt{s}$ for each value of $m_\Phi^{}
 (=m_H^{}=m_A^{}=m_{H^\pm}^{})$ with $\sin(\beta-\alpha)=1$,
 $\tan\beta=1$ and $M=0$. 
 The case for $m_h=120$ [160] GeV is shown in the left [right] figure. 
}
\label{figure3}
\end{center}
\end{figure}

In Fig.~\ref{figure3}, the full cross sections of $e^-e^- \to
\gamma\gamma \to hh$ 
are shown for $m_h=120$ GeV in the left figure and $m_h=160$ GeV in the
right figure, respectively, 
as a function of $\sqrt{s}$ for various values of the extra Higgs boson
masses $m_{\Phi}^{}$ ($\equiv m_{H}^{}=m_A^{}=m_{H^\pm}^{}$)
in the cases of $\tan\beta=1$, $\sin(\beta-\alpha)=1$ and $M = 0$.
In order to extract the contribution from $\hat{\sigma}(+,+)$ that is
 sensitive to the $hhh$ vertex, we take the polarizations of the initial laser beam
to be both $-1$, and those for the initial electrons to be both +0.45 \cite{jikia}.
The full cross section for $m_{\Phi}^{}=400$ GeV has similar energy
dependences to the sub cross section $\hat{\sigma}(+,+)$ 
in Fig.~\ref{figure2}, where corresponding energies are rescaled
approximately by around $\sqrt{s} \sim E_{\gamma\gamma}/0.8$ due to
the photon luminosity spectrum. 
For smaller $m_{\Phi}^{}$, the peak around $\sqrt{s} \sim 350$ GeV becomes
lower because of smaller $\Delta \Gamma_{hhh}^{\rm THDM}/\Gamma_{hhh}^{\rm SM}$.

\begin{figure}[t]
\begin{center}
\includegraphics[width=7.8cm,angle=-90]{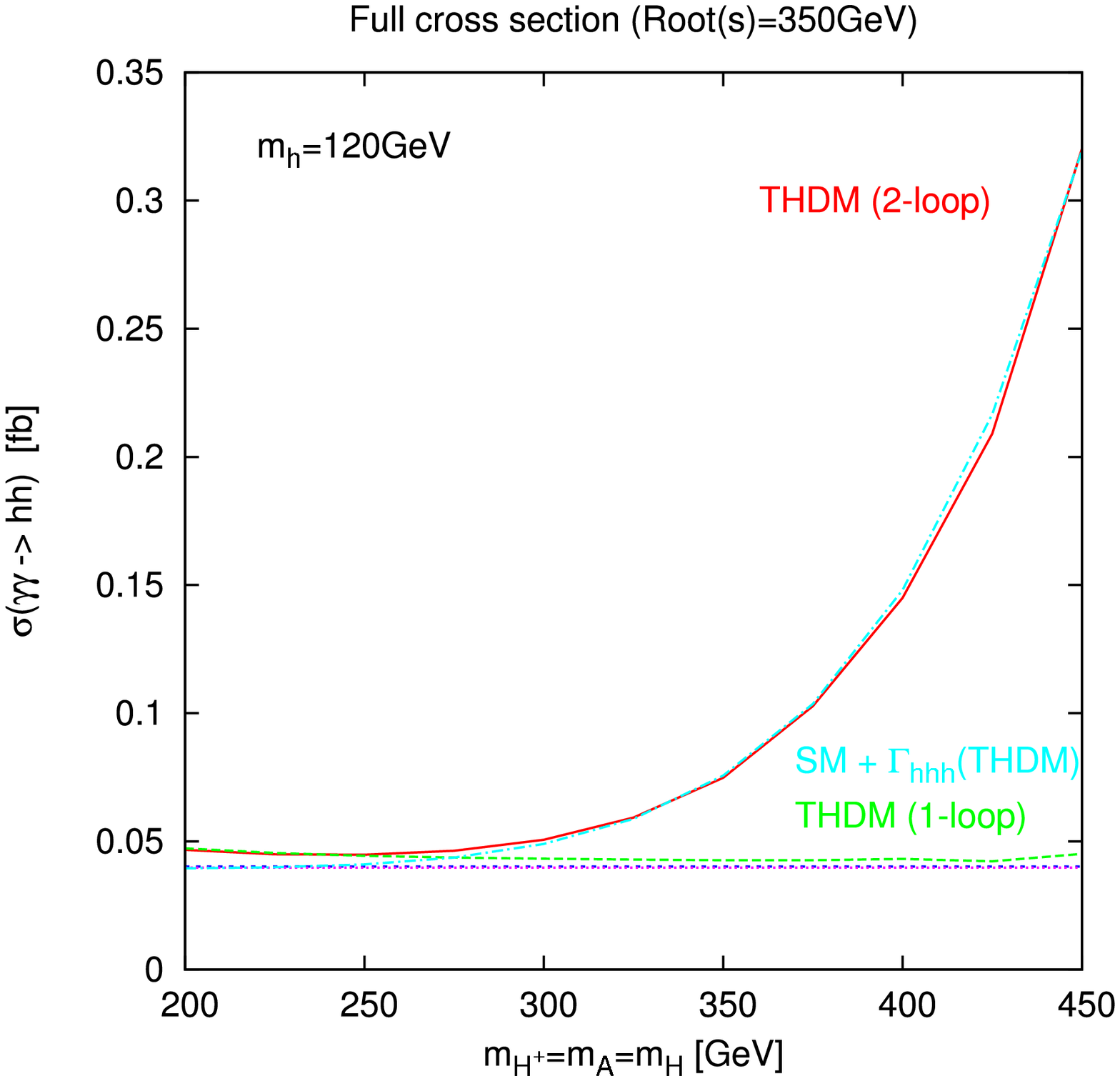}
\includegraphics[width=7.8cm,angle=-90]{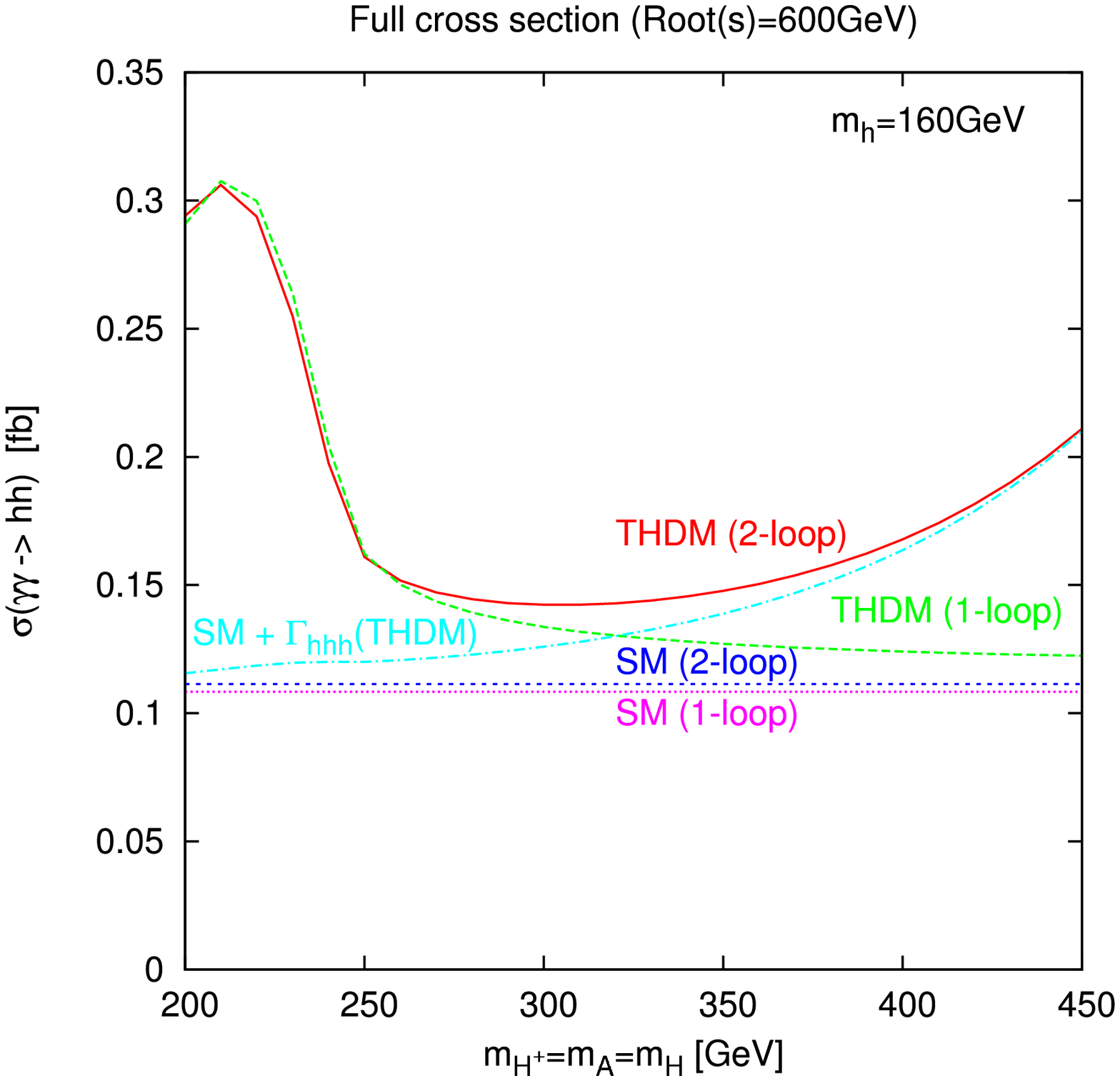}
\caption{%
  In the left [right] figure, the full cross section of $e^-e^- \to \gamma\gamma \to hh$ 
 at $\sqrt{s} = 350$ GeV [600 GeV] for $m_h=120$ [160] GeV is shown as a
 function of $m_\Phi^{} (=m_H^{}=m_A^{}=m_{H^\pm}^{})$ with 
 $\sin(\beta-\alpha)=1$, $\tan\beta=1$ and $M=0$.
}
\label{figure4}
\end{center}
\end{figure}
In Fig.~\ref{figure4},  the full cross sections are shown as a 
function of $m_{\Phi}^{}$ for $m_h=120$ GeV at $\sqrt{s}=350$ GeV (the
left figure) and $m_h=160$ GeV at $\sqrt{s}=600$ GeV (the right figure). 
In each figure, five curves correspond to the cases (a) to (e) in
Fig.~2. The other parameters are taken to be $\sin(\beta-\alpha)=1$,
$\tan\beta=1$ and $M=0$.  
In the left figure,
one can see that the cross section is enhanced due to the enlarged
$\Gamma_{hhh}^{\rm THDM}$ for larger values of $m_{\Phi}^{}$    
which is proportional to $m_{\Phi}^4$ (when $M \sim 0$).
This implies that the cross section for these parameters 
is essentially determined by the pole diagram contributions.
 The effect of the charged Higgs boson loop from $\Delta {\cal
M}$ is relatively small since
the threshold of charged Higgs boson production is far.  
Therefore, the deviation in the cross section from the SM value
 is smaller for relatively small $m_{\Phi}^{}$ (10-20\% for $m_{\Phi} <
 300$ GeV due to the charged Higgs loop effect in $\Delta {\cal
M}$) but it becomes
 rapidly enhanced for greater values of $m_{\Phi}^{}$ (${\cal O}(100)$
 \% for $m_{\Phi} > 350$ GeV due to the large $\Delta \Gamma_{hhh}^{\rm THDM}$).
A similar enhancement for the large $m_{\Phi}$ values can be seen
in the right figure.
The enhancement in the cross section in the THDM
can also be seen for
$m_\Phi^{} < 250$ GeV, where the threshold effect of the
charged Higgs boson loop in $\Delta {\cal M}$ appears around $\sqrt{s} \sim 600$ GeV in
addition to that of the top quark loop diagrams in ${\cal M}$. 
For $m_\Phi^{}=250$-$400$ GeV, both contributions from 
the charged Higgs boson loop contribution and the
effective $hhh$ coupling are important
and enhance the cross section from its SM value by 40-50\%.


We have analysed the new physics loop effects on the cross section
of $\gamma\gamma \to hh$ in the THDM including the next to leading
effect due to the extra Higgs boson loop diagram in the $hhh$ vertex.
Our analysis shows that the cross section can be largely changed from the SM prediction by the two kinds
of contributions; i.e., additonal contribution by the charged
Higgs boson loop in $\Delta {\cal M}$,
and the effective one-loop $hhh$ vertex $\Gamma_{hhh}^{\rm THDM}$
enhanced by the non-decoupling effect of extra Higgs bosons.
The cross section strongly depends on $m_h$ and $\sqrt{s}$
and also on $m_{\Phi}^{}$.
The approximation of the full cross section in the case (a) (2-loop
THDM) by using the result in the case (e)
(SM+$\Gamma_{hhh}^{\rm THDM}$) is a good description for  $\sqrt{s} \ll 2
m_\Phi/0.8$.
On the other hand, in a wide region between threshold of top pair
production and that of charged Higgs boson pair production,
both the contributions (those from $\Delta {\cal M}$ and
from $\Gamma_{hhh}^{\rm THDM}$) are important. 
In the region below the threshold of the real production of
extra Higgs bosons, the cross section can be a few times 0.1 fb
in the THDM while that in the SM is about 0.05 fb.
Such differences from the SM prediction would be detectable at a future
photon collider.

We note that the analysis in this letter can be applied to the models \cite{Zee}
in which extra charged scalar bosons appear with a potentially large
loop correction in the $hhh$ coupling. 

The work of S. K. was supported in part by Grant-in-Aid for Science
Research, Japan Society for the Promotion of Science (JSPS),
No. 18034004. The work of Y. O. was supported in part by Grant-in-Aid for 
Science Research, MEXT-Japan, No. 16081211, and JSPS, No. 20244037.\\

{\it Note added:}
After this work was finished, we noticed the paper \cite{hollik} 
which studied $\gamma\gamma\to hh$ in the THDM.
Our paper includes the additional contribution of the $hhh$ vertex
(the leading two-loop effect on $\gamma\gamma\to hh$), 
which was not considered in \cite{hollik}.


\section*{Appendix}

If the mass of the particle in the loop comes from the VEV, a large mass
implies a large coupling constant, so that a naive argument
of the decoupling theorem is not applied. It is known that in such a
case a powerlike mass contribution of particles in the loop appears in
the one-loop contribution. This is called the non-decoupling effect. 

When one-loop corrected $hhh$ vertex $\Gamma_{hhh}^{\rm THDM}$ largely
deviates from $\Gamma_{hhh}^{\rm SM}$ due to the non-decoupling
property of the extra Higgs bosons,
the main two loop contribution to $\gamma\gamma\to hh$ comes from the
$s$-channel diagrams with the effective $hhh$ coupling. 
We here show this by the use of a power counting method.
For simplicity, we consider the leading powerlike effect of the mass
of particles in the loops in the two-loop diagrams for the case with
$M \sim 0$ where masses of extra Higgs bosons are proportional to
the VEV so that the non-decoupling effect is maximal.

\begin{figure}[t]
\begin{center}
\includegraphics[width=12cm,angle=0]{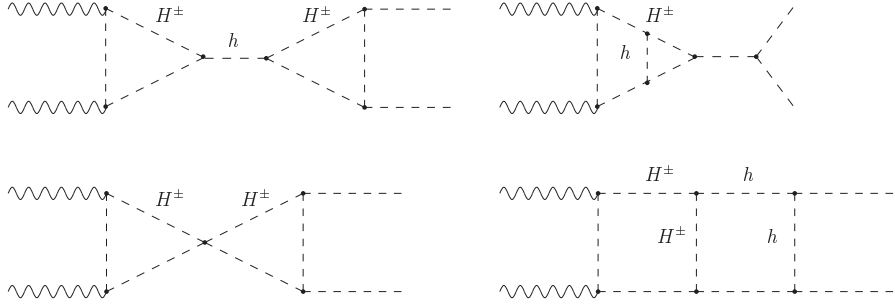}
\caption{%
Example of the two-loop diagrams contributing to $\gamma\gamma \to hh$.
}
\label{figure5}
\end{center}
\end{figure}
When $M\sim 0$, the coupling constants of $hH^+ H^-$ and $hhH^+ H^-$ are
proportional to $m_{H^\pm}^{2}/v$ and $m_{H^\pm}^{2}/v^2$, respectively.
We consider the situation that $m_{H^\pm}^{} \gg \sqrt{s} > 2 m_h$.
The leading non-decoupling effect of the $H^\pm$ one-loop triangle-type diagram
in Fig.~\ref{figure}(up-right) and that of the $H^\pm$ one-loop box-type diagram 
in Fig.~\ref{figure}(bottom-left) are evaluated as
\begin{eqnarray}
&& {\cal M}^{\rm 1-loop}_{\rm trig} \propto \frac{1}{16\pi^2}
  \frac{q^2}{v}\frac{1}{s-m_h^2}\left(\frac{m_h^2}{v}\right)
   \sim \frac{q^2}{(4\pi v)^2}, \\
&& {\cal M}^{\rm 1-loop}_{\rm box} \propto 
  \frac{1}{16\pi^2} \frac{q^2}{v^2}
   \sim \frac{q^2}{(4\pi v)^2}, 
\end{eqnarray}
where we used the fact that the effective $\gamma\gamma h$ and
$\gamma\gamma hh$ vertices come
from the dimension six operator $|\Phi_i|^2 F_{\mu\nu}F^{\mu\nu}$, so
that they are proportial to $q^2/v$ and $q^2/v^2$  at
the leading order, respectively, where $q^2 \sim s$.
Therefore, the effect of $m_{H^\pm}$ on $\gamma\gamma\to hh$ can be at
most $\log m_{H^\pm}^{}$ at the one-loop level.
A similar conclusion of power counting can also be obtained for 
one-loop effects of top and bottom quarks and $W$ bosons to $\gamma\gamma\to hh$.  

Next, let us examine two-loop diagrams shown in Fig.~\ref{figure5}.
The non-decoupling effect in the diagram (a) in
Fig.~\ref{figure5}(up-left) is calculated as
\begin{eqnarray}
 {\cal M}^{\rm 2-loop}_{\rm (a)} \propto \left(\frac{1}{16\pi^2}\right)^2
  \frac{q^2}{v} \frac{1}{s-m_h^2}
  \left(\frac{m_{H^\pm}^2}{v}\right)^3
  \frac{d^4 k}{(k^2-m_{H^\pm}^2)^3}
 \sim \frac{q^2}{(4\pi v)^2}
      \left(\frac{m_{H^\pm}^4}{(4\pi v)^2 m_h^2}\right), 
 \end{eqnarray}
where momenta of external lines are neglected, and
$k$ is the momentum in the loop of the effective $hhh$ vertex,
which is replaced by the greatest dimensionful parameter of the system; i.e. $m_{H^\pm}^{}$.
This result of the power counting is not changed even after the renormalization of
the $hhh$ vertex is performed\cite{hhh-thdm1}. 
There are other two loop diagrams which are generated from the s-channel
type one-loop diagram, such as the diagram (b) in
Fig.~\ref{figure5}(up-right) where there is the bridge of $h$ in the
 $H^\pm$ triangle type loop.
Its non-decoupling effect is evaluated as 
\begin{eqnarray}
 {\cal M}^{\rm 2-loop}_{\rm (b)} \propto \left(\frac{1}{16\pi^2}\right)^2
  \frac{q^2}{v} \left(\frac{m_{H^\pm}^2}{v}\right)^2
    \frac{d^4k}{(k^2-m_{H^\pm}^2)^3}
\frac{1}{s-m_h^2}
  \left(\frac{m_{h}^2}{v}\right)^2
 \sim \frac{q^2}{(4\pi v)^2}
      \left(\frac{m_{H^\pm}^2}{(4\pi v)^2}\right). 
 \end{eqnarray}
The dependence on $m_{m_H^\pm}^{}$ is not quartic but quadratic.  
We have examined all the other two-loop diagrams which are generated
from the one-loop $s$-channel diagram and confirmed that they are
the same or less power dependence on $m_H^\pm$ as the diagram (b). 

A similar counting can also be applied for the diagrams such as
the diagram (c) in Fig.~\ref{figure5}(down-left) where charged Higgs bosons are
running in the both loops, and the diagram
(d) in Fig.~\ref{figure5}(down-right) where ladder of
$h$ is added to the one-loop box type diagram; 
\begin{eqnarray}
 &&
 {\cal M}^{\rm 2-loop}_{\rm (c)} \propto \left(\frac{1}{16\pi^2}\right)^2
  \frac{q^2}{v^2} 
    \frac{d^4k}{(k^2-m_{H^\pm}^2)^3}
  \left(\frac{m_{H^\pm}^2}{v}\right)^2
 \sim \frac{q^2}{(4\pi v)^2}
      \left(\frac{m_{H^\pm}^2}{(4\pi v)^2}\right),\\ 
&& M^{\rm 2-loop}_{\rm (d)} \propto \left(\frac{1}{16\pi^2}\right)^2
  \frac{q^2}{v^2} 
    \frac{d^4k}{(k^2-m_{h}^2)^3}
  \left(\frac{m_{h}^2}{v}\right)^2
 \sim \frac{q^2}{(4\pi v)^2}
      \left(\frac{m_h^4}{(4\pi v)^2 m_{H^\pm}^2}\right).
 \end{eqnarray}
We find that all the 1PI two-loop diagrams of $\gamma\gamma hh$
also have the quadratic or less power dependences on $m_{H^\pm}$. 

The power dependence on $m_{H^\pm}$ in the two point function of $h$
can be reduced by the renormalization of mass $m_h^2$, but 
the highest power of $m_{H^\pm}$ in the 1PI two loop
diagrams of $\gamma\gamma hh$ does not change by the renormalization.

In conclusion, the non-decoupling effect of $H^\pm$ on the renormalized
amplitude of $\gamma\gamma \to hh$ at the two loop level can be
described as
\begin{eqnarray}
{\cal M}^{\rm 2-loop} \propto \frac{q^2}{(4\pi v)^2} \left[1 + 
      {\cal O}\left(\frac{m_{H^\pm}^4}{(4\pi v)^2 m_h^2}\right) + 
      {\cal O}\left(\frac{m_{H^\pm}^2}{(4\pi v)^2}\right)
                                       \right], 
\end{eqnarray}
where the second term in RHS comes from the $s$-channel diagrams which include
the one-loop corrected $hhh$ vertex. For the case where non-decoupling
property of the extra Higgs bosons is important, the contribution from
this term is dominant when $m_{H^\pm}^{} \gg m_h$.
Although we gave the explanation for  the charged Higgs loop
effects, this argument can also be applied to loop effects
of all quarks, gauge bosons and extra Higgs bosons with non-decoupling property.

\end{document}